\documentclass[reprint,showpacs,amsmath,amssymb,aps,pre]{revtex4-2}
\usepackage{dcolumn}
\usepackage{bm}
\usepackage[dvipdfmx]{graphicx}
\usepackage[dvipdfmx]{hyperref}
\usepackage[mathlines]{lineno}
\usepackage{physics}

\begin{document}

\title{Wilson-loop formulations of the spin polarization and the spin Chern number}
\author{Kiminori Hattori and Ryuki Uchida}
\affiliation{Department of Systems Innovation, Graduate School of Engineering Science, The University of Osaka, Toyonaka, Osaka 560-8531, Japan}
\date{\today}

\begin{abstract}	
In this paper, we derive the spin polarization and the spin Chern number in momentum and real spaces in terms of the Wilson loop in each spin sector.
For these derivations, the spin matrix in the subspace of occupied or unoccupied states is generalized by the unitary rotation to prevent the spectral gap from closing.
In the presence of a finite spin gap, the spin eigenvectors recompose the eigenstates of the Hamiltonian in two spin sectors.
A dimensional reduction leads to the charge polarization in each spin sector in terms of the Wilson loop along the effective Brillouin zone (BZ) in one dimension (1D).
A linear combination of the charge polarizations defines the spin polarization, which naturally quantifies the spin Chern number in momentum space.
It is also shown that the spin Chern number is directly derived from the local Wilson loops on a discrete 2D BZ.
The theoretical treatment is further generalized for real space in a supercell framework, which can deal with ordered and disordered systems in a unified manner.
In this framework, the spin Chern number in real space is formulated in terms of the Wilson loop around the BZ boundary.
These formulations are used to numerically evaluate topological phases in the Bernevig-Hughes-Zhang model.
Numerical results show an equivalence of the spin Chern number and the ${\mathbb{Z}_2}$ index.
\end{abstract}

\maketitle

\section{Introduction}
\label{sec:1}

The discovery of topological insulators with time-reversal symmetry in two dimensions (2D) has initiated the widespread study of topological aspects in insulators, semimetals, and superconductors \cite{ref:1,ref:2}.
Such an exotic 2D system is originally known as the quantum spin Hall insulator \cite{ref:3,ref:4}.
The quantum spin Hall state consists of two copies of quantum Hall states with opposite Chern numbers, arranged so that time-reversal symmetry is preserved.
In a strip of the 2D insulator, there exist helical edge states at two lateral boundaries due to the bulk-boundary correspondence.
These 1D states are essentially insusceptible to time-reversal invariant perturbations.
Parallel to these studies, a classification scheme predicting topological invariants definable for given systems was established \cite{ref:5,ref:6}.
This scheme depends only on spatial dimensions and underlying discrete symmetries, including time-reversal symmetry, particle-hole symmetry, and a combination of the two, known as chiral symmetry.
The time-reversal symmetric topological insulators belong to the symplectic class AII, which allows a ${\mathbb{Z}_2}$ invariant in 2 or 3D.
The topological invariant identifies if a system of interest is topologically trivial or not.
Therefore, its formulation is crucial to the study of topological matter.

Recently, a number of approaches in this direction have been proposed, such as the ${\mathbb{Z}_2}$ index based on the time-reversal polarization \cite{ref:7,ref:8}, the spin-dependent twisted boundary phases \cite{ref:9,ref:10}, the Berry connection and curvature in half of the Brillouin zone (BZ) \cite{ref:8,ref:11,ref:12}, the evolution of Wannier states during time-reversal pumping \cite{ref:13,ref:14}, the artificial flux insertion \cite{ref:15}, the gauge independent ${\mathbb{Z}_2}$ index \cite{ref:16,ref:17}, the Pfaffian-Bott index \cite{ref:18,ref:19,ref:20,ref:21}, and the scattering matrix approach \cite{ref:22,ref:23}.
It is also natural to consider the spin operator in decomposing time-reversed pairs of eigenstates, since it is odd under time reversal.
Diagonalizing the spin operator projected onto occupied eigenstates defines the projector onto each spin sector.
The spin projection leads to the spin Chern number \cite{ref:24,ref:25}, the spin Bott index \cite{ref:26,ref:27}, the single-point spin Chern number \cite{ref:28}, and the local spin Chern marker \cite{ref:29,ref:30}.

The topological invariants are defined mostly in momentum space for ordered systems with translational symmetry.
Conventional formulations in momentum space are, however, insufficient in elucidating the topology of inhomogeneous or disordered systems that lack translational invariance.
In this sense, it is desirable to propose a unified theory that enables momentum and real-space formulations and deals with ordered and disordered systems on an equal footing.
In this paper, we derive the spin polarization and the spin Chern number $\nu$ from the Wilson loop in each spin sector.
The ${\mathbb{Z}_2}$ index is given by $\nu$ modulo 2.
The invariant $\nu$ is gauge independent, numerically quantified, formulated in momentum and real spaces, and becomes a useful equivalent to the previous ones.

This paper is organized as follows.
In Sec. \ref{sec:2}, we briefly review the Bernevig-Hughes-Zhang (BHZ) model \cite{ref:4,ref:31}.
A spin mixing term is incorporated into this model.
The generalized BHZ model is used in benchmarking the present formulations.
In Sec. \ref{sec:3}, we formulate the charge polarization in each spin sector in terms of the Wilson loop along an effective BZ in 1D.
A linear combination of the charge polarizations defines the spin polarization, which naturally gives the spin Chern number in momentum space.
In Sec. \ref{sec:4}, we show that the spin Chern number is directly derived from the local Wilson loops on a discrete 2D BZ.
In Sec. \ref{sec:5}, we generalize the theoretical treatment for a real-space formulation in a supercell framework, which can deal with ordered and disordered systems in a unified manner.
In this framework, the spin Chern number in real space is formulated in terms of the Wilson loop around the BZ boundary.
Finally, Sec. \ref{sec:6} provides the summary of our research.

\section{Model}
\label{sec:2}

We begin by considering the Qi-Wu-Zhang (QWZ) model \cite{ref:31,ref:32}.
This model is described by the 2D lattice Hamiltonian
\begin{equation}
\label{eq:1}
\begin{split}
H_0 &= m \sum\limits_{l_x, l_y } {\ketbra{l_x, l_y }{l_x, l_y } \otimes {\tau _z}}  \\ 
&+ t\sum\limits_{l_x, l_y } {(\ketbra{l_x +1, l_y }{l_x, l_y } \otimes \frac{{{\tau _z} + i{\tau _x}}}{2} + {\text{H.c.}})}  \\ 
&+ t\sum\limits_{l_x, l_y } {(\ketbra{l_x, l_y +1}{l_x, l_y } \otimes \frac{{{\tau _z} + i{\tau _y}}}{2} + {\text{H.c.}})} , \\ 
\end{split}
\end{equation}
where $\ket{{l_x},{l_y}}$ represents the basis ket at each site labeled with ${l_\mu } \in \{1,2, \ldots ,{N_\mu }\} $ for $\mu \in \{x,y\} $, $t$ denotes the intersite hopping energy, and $m$ refers to the orbital energy difference in each site.
The Pauli matrices $\boldsymbol{\tau}= ({\tau _x},{\tau _y},{\tau _z})$ act on the orbital degrees of freedom.
Applying double Fourier transformation to the 2D model in a torus geometry, the two-band Hamiltonian is formulated as
\begin{equation}
\label{eq:2}
{H_0}({\mathbf{k}}) = {\mathbf{h}}({\mathbf{k}}) \cdot \boldsymbol{\tau}
\end{equation}
in momentum space, where ${\mathbf{k}} = ({k_x},{k_y})$.
The vector ${\mathbf{h}} = ({h_x},{h_y},{h_z})$ is composed of ${h_x} = t \sin{k_x}$, ${h_y} = t \sin{k_y}$, and ${h_z} = m + t (\cos{k_x} + \cos{k_y})$.
In this model, time-reversal symmetry is broken.
Following a previous related study \cite{ref:33}, we find that the topologically nontrivial phases with nonzero Chern numbers $C = \pm 1$ emerge in the region where $0 < {m^2} < 4{t^2}$.

The two-band QWZ model is a building block for the four-band BHZ model that respects time-reversal symmetry \cite{ref:31}.
The latter is formulated simply as
\begin{equation}
\label{eq:3}
{\mathcal{H}_0}({\mathbf{k}}) = \frac{{{s_0}+{s_z}}}{2} \otimes {H_0}({\mathbf{k}}) + \frac{{{s_0}-{s_z}}}{2} \otimes H_0^{*} (-{\mathbf{k}}) ,
\end{equation}
where the Pauli matrices $\boldsymbol{s} = ({s_x},{s_y},{s_z})$ deal with the spin degrees of freedom and ${s_0}$ denotes the $2 \times 2$ identity matrix.
The Hamiltonian ${\mathcal{H}_0}$ conserves the spin ${s_z}$.
Since ${\mathcal{H}_0}$ also preserves spatial inversion symmetry, the ${\mathbb{Z}_2}$ invariant, denoted as ${\nu _{\mathbb{Z}_2}}$ in the following, is derived from the parity eigenvalues of Kramers degenerate pairs in occupied bands \cite{ref:34}.
It is easily found that ${\nu _{\mathbb{Z}_2}} = 1$ for ${m^2} < 4{t^2}$ and it is 0 otherwise.
These criteria do not contradict the ones for $C$ of the QWZ model.
In addition, we consider a spin mixing term
\begin{equation}
\label{eq:4}
{\mathcal{H}_1}({\mathbf{k}}) = {s_x} \otimes g{\tau _y} ,
\end{equation}
where $g$ denotes the strength of spin coupling \cite{ref:31}.
The Hamiltonian ${\mathcal{H}_1}$ preserves time-reversal symmetry but violates inversion symmetry as well as spin conservation.
A generalized BHZ Hamiltonian consists of $\mathcal{H} = {\mathcal{H}_0} + {\mathcal{H}_1}$.

\section{Spin polarization in momentum space}
\label{sec:3}

Let the eigenequation in momentum space be $\mathcal{H}({\mathbf{k}}) \ket{{u_n}({\mathbf{k}})} = {\varepsilon _n} \ket{{u_n}({\mathbf{k}})} $.
Before describing the spin polarization, we first introduce the spin matrix for occupied (or unoccupied) eigenstates.
A natural choice is ${S_{nn'}}({\mathbf{k}}) = \mel {{u_n}({\mathbf{k}})} {s_z} {{u_{n'}}({\mathbf{k}})} $.
This matrix can be expressed as $S({\mathbf{k}}) = {U^\dag }({\mathbf{k}}){s_z}U({\mathbf{k}})$, where ${U_{\alpha n}}({\mathbf{k}}) = \braket {\alpha }{{u_n}({\mathbf{k}})} $ and $\alpha $ represents the internal degrees of freedom.
The spin matrix is generalizable by adopting ${s_{\boldsymbol \theta}} = {e^{i{\boldsymbol \theta} \cdot \boldsymbol{s}/2}}{s_z}{e^{-i{\boldsymbol \theta} \cdot \boldsymbol{s}/2}}$ instead of ${s_z}$.
In this study, we use such an extended form
\begin{equation}
\label{eq:5}
S({\mathbf{k}}) = {U^\dag }({\mathbf{k}}){s_\theta }U({\mathbf{k}})
\end{equation}
for $\boldsymbol \theta = \theta {{\mathbf{e}}_y}$, where ${s_\theta } = {e^{i\theta {s_y}/2}}{s_z}{e^{-i\theta {s_y}/2}}$ and ${{\mathbf{e}}_y}$ denotes the unit vector in the $y$ direction.
Clearly, ${s_\theta} = {s_z}$ for $\theta = 0$.
Note that the unitary rotation is restricted to $S({\mathbf{k}})$ and does not alter the spin operators contained in $\mathcal{H}({\mathbf{k}})$.

Solving the eigenequation $S({\mathbf{k}}) \ket{{v_i}({\mathbf{k}})} = {s_i}({\mathbf{k}}) \ket{{v_i}({\mathbf{k}})}$, one derives the spin eigenvalues ${s_i}({\mathbf{k}})$ and the spin eigenvectors $\ket{{v_i}({\mathbf{k}})}$.
The positive and negative signs of ${s_i}({\mathbf{k}})$ distinguish two spin sectors unless the spectral gap vanishes.
In the following, the spin eigenvector $\ket{{v_{i\sigma}}({\mathbf{k}})}$ in each spin sector is denoted by $\sigma = \operatorname{sgn} ({s_i})$.
The spin eigenvector is expanded in terms of the energy eigenstates as
\begin{equation}
\label{eq:6}
\ket {{u_{i\sigma }}({\mathbf{k}})} = \sum\limits_n {\ket{{u_n}({\mathbf{k}})} \hspace{-3pt} \braket{{u_n}({\mathbf{k}})}{{v_{i\sigma}}({\mathbf{k}})}} .
\end{equation}
In matrix notation, this relation is expressed as ${U_\sigma }({\mathbf{k}}) = U({\mathbf{k}}){V_\sigma }({\mathbf{k}})$, where $[{U_{\sigma }}({\mathbf{k}})]_{\alpha i} = \braket {\alpha }{{u_{i\sigma }}({\mathbf{k}})} $ and $[{V_\sigma }({\mathbf{k}})]_{ni} = \braket{{u_n}({\mathbf{k}})}{v_{i\sigma}({\mathbf{k}})}$.
Note that Eq. (\ref{eq:6}) recomposes the eigenstates of the Hamiltonian in two spin sectors.

Following the arguments for the time-reversal polarization \cite{ref:8} and the subsequent reformulation in terms of the Wannier states \cite{ref:13,ref:14}, we regard ${k_x}$ as a cyclical time variable and assume a fictitious 1D periodic system parallel to the $y$ axis.
Here, we suppose the lattice points ${k_\ell } = \tfrac{{2\pi \ell }}{M}$ and discretize the 1D BZ, where $\ell \in \{ 0,1, \ldots ,M - 1\} $.
The overlap matrix between two adjacent points is written as $q_\sigma ^{(\ell + 1,\ell )}({k_x}) = U_\sigma ^\dag ({k_x},{k_{\ell + 1}}){U_\sigma }({k_x},{k_\ell })$.
The path-ordered product of overlap matrices
\begin{equation}
\label{eq:7}
{Q_\sigma }({k_x}) = q_\sigma ^{(0,M-1)}({k_x}) \cdots q_\sigma ^{(2,1)}({k_x})q_\sigma ^{(1,0)}({k_x})
\end{equation}
forms a Wilson loop.
In the limit of $M \to \infty $, $q_\sigma ^{(\ell + 1,\ell )}({k_x}) = {e^{i({k_{\ell + 1}} - {k_\ell }){{\mathbf{e}}_y} \cdot {{\mathbf{A}}_\sigma }({k_x},{k_\ell })}}$ so that
\begin{equation}
\label{eq:8}
{Q_\sigma }({k_x}) = \mathcal{P}{e^{i\oint_{\mathcal{C}({k_x})} {d{\mathbf{k}} \cdot {{\mathbf{A}}_\sigma }({\mathbf{k}})} }} ,
\end{equation}
where $\mathcal{P}$ denotes path ordering, $\mathcal{C}({k_x})$ stands for the circumference of the effective 1D BZ for a fixed ${k_x}$, and ${{\mathbf{A}}_\sigma }({\mathbf{k}}) = iU_\sigma ^\dag ({\mathbf{k}}){\nabla _{\mathbf{k}}}{U_\sigma }({\mathbf{k}})$ is the Berry connection.
Thus, the charge polarization is formulated in each spin sector as
\begin{equation}
\label{eq:9}
\begin{split}
{P_\sigma }({k_x}) &= \frac{1}{{2\pi }} \Tr\oint_{\mathcal{C}({k_x})} {d{\mathbf{k}} \cdot {{\mathbf{A}}_\sigma }({\mathbf{k}})} \\
 &= \frac{1}{{2\pi }}\sum\limits_p {\arg {\lambda _{p\sigma }}({k_x})} , \\
\end{split}
\end{equation}
where ${\lambda _{p\sigma }}({k_x})$ is the eigenvalue of ${Q_\sigma }({k_x})$.

Note that the pumped charge per cycle is quantized to be ${P_\sigma }(2\pi ) - {P_\sigma }(0) = {C_\sigma }$, where
\begin{equation}
\label{eq:10}
{C_\sigma } = \frac{1}{{2\pi }}\Tr\oint_{\partial {\text{BZ}}} {d{\mathbf{k}} \cdot {{\mathbf{A}}_\sigma }({\mathbf{k}})}
\end{equation}
is the Chern number in each spin sector \cite{ref:31,ref:35,ref:36}.
Since ${k_x} = \pi $ is a time-reversal symmetric point, it is verified that ${P_\sigma }(2\pi ) - {P_\sigma }(\pi ) = {P_\sigma }(\pi ) - {P_\sigma }(0) = {C_\sigma }/2$.
Hence, the spin Chern number, defined by
\begin{equation}
\label{eq:11}
\nu = \frac{{{C_ + } - {C_ - }}}{2} ,
\end{equation}
can be rewritten as
\begin{equation}
\label{eq:12}
\nu = {P_s}(\pi ) - {P_s}(0) ,
\end{equation}
where ${P_s}({k_x}) = {P_ + }({k_x}) - {P_ - }({k_x})$ is the spin polarization.
The spin polarization is an analog to the time-reversal polarization.
Time-reversal invariance requires that ${C_ + } + {C_ - } = 0$.
As a result, $\nu = {C_ + } = - {C_ - }$ is quantized to be an integer.
The invariant ${\nu _{{\mathbb{Z}_2}}} \in \{0,1\} $ is given by $\nu$ modulo 2.
The above argument is generically valid for time-reversal invariant 2D systems.
In the case of the BHZ model, ${Q_\sigma }({k_x})$ is a scalar and thereby ${P_\sigma }({k_x}) = \tfrac{1}{{2\pi }} \arg {Q_\sigma }({k_x})$.
It may be worth noting that the spin Chern number argued here is explicitly gauge invariant.
This feature is absent in the original ${\mathbb{Z}_2}$ index derived from the time-reversal polarization \cite{ref:7,ref:8}.
The charge polarizations ${P_ \pm }({k_x})$ are equivalent to the center positions of the Wannier states in two spin sectors.
It should be emphasized here that Eq. (\ref{eq:12}) numerically quantifies the relevant invariant $\nu$.
This feature is absent in the original Wannier function formalism \cite{ref:13,ref:14}.

\begin{figure}
\centering
\includegraphics{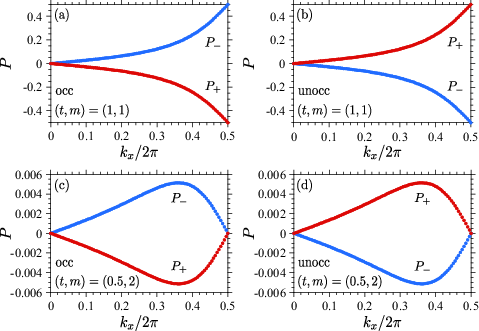}
\caption{Charge polarization ${P_\sigma }$ as a function of ${k_x}$ for [(a) and (b)] $(t,m) = (1,1)$ and [(c) and (d)] $(t,m) = (0.5,2)$. The left panels, (a) and (c), and the right panels, (b) and (d), show the numerical results for occupied and unoccupied states, respectively. The parameters used in the calculation are $g = \theta = 0$ and $M = 100$. `occ' and `unocc' denote the occupied and unoccupied states, respectively.}
\label{fig:1}
\end{figure}

\begin{figure}
\centering
\includegraphics{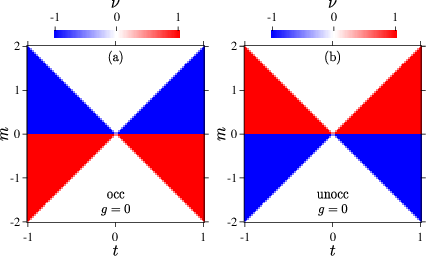}
\caption{Phase diagrams of momentum-space spin Chern number $\nu $ in the parameter space $(t,m)$ for (a) occupied and (b) unoccupied states. The parameters used in the calculation are $g = \theta = 0$ and $M = 100$. `occ' and `unocc' denote the occupied and unoccupied states, respectively.}
\label{fig:2}
\end{figure}

Figure \ref{fig:1} shows the charge polarizations ${P_\sigma }$ as a function of ${k_x}$ for occupied and unoccupied states.
These numerical results are derived for the BHZ model by assuming $g = \theta = 0$.
In the case of $(t,m) = (1,1)$, ${P_\sigma }(\pi ) - {P_\sigma }(0) = - \sigma /2$ for occupied states.
This indicates that ${C_\sigma } = - \sigma $ and hence $\nu = - 1$.
It is also found that ${C_\sigma } = \sigma $ and $\nu = 1$ for unoccupied states.
The polarization spectra imply that two Wannier states with opposite spins shift in the opposite directions.
These states are Kramers degenerate at ${k_x} = 0$ because of time-reversal symmetry.
At ${k_x} = \pi $, their center positions are separated by 1 (i.e., one lattice constant).
Then, Kramers pairs switch partners and time-reversal symmetry is recovered \cite{ref:8}.
In the case of $(t,m) = (0.5,2)$, ${P_\sigma }(\pi ) = {P_\sigma }(0)$ so that ${C_\sigma } = \nu = 0$ for both occupied and unoccupied states.
In this case, the two Wannier states eventually meet at ${k_x} = \pi $.
Figure \ref{fig:2} summarizes the phase diagrams of $\nu$ in the parameter space $(t,m)$ derived from the spin polarization for $g = \theta = 0$.
As is clearly seen in the figure, the spin Chern number is quantized to be $\nu = \pm 1$ in the topologically nontrivial regime where $0 < {m^2} < 4{t^2}$ and it is 0 otherwise.
In this figure, one finds a phase boundary at $m = 0$, where $\nu $ changes by $\pm 2$.
Note that no phase transition occurs there in view of ${\nu _{\mathbb{Z}_2}}$ defined modulo 2.

\begin{figure}
\centering
\includegraphics{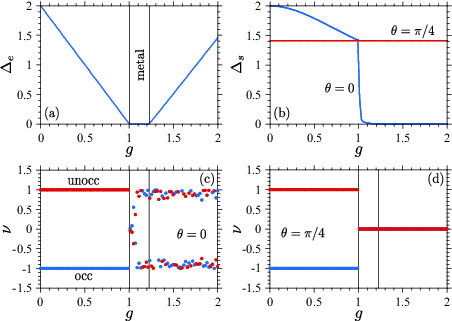}
\caption{(a) Energy gap ${\Delta _e}$ and (b) spin gap ${\Delta _s}$ as a function of $g$. In panel (b), the numerical results are compared for $\theta = 0$ and $\tfrac{\pi }{4}$. The lower two panels show momentum-space spin Chern number $\nu $ as a function of $g$ for (c) $\theta = 0$ and (d) $\theta = \tfrac{\pi }{4}$. The parameters used in the calculation are $(t,m) = (1,1)$ and $M = 100$. The thin lines inserted in the figure indicate the phase boundaries expected from the energy-gap closure. `occ' and `unocc' denote the occupied and unoccupied states, respectively.}
\label{fig:3}
\end{figure}

Figures \ref{fig:3} (a) and \ref{fig:3} (b) show how the energy and spin gaps vary with $g$, respectively.
In these calculations, we assumed $(t,m) = (1,1)$.
The energy gap ${\Delta _e}$ vanishes in a finite range where $1 \leq g \leq 1.2$.
In this region, the system becomes metallic.
The spin gap ${\Delta _s}$ tends to vanish in the region where $g \geq 1$ for $\theta = 0$, whereas ${\Delta _s}$ remains finite in the entire region for $\theta = \tfrac{\pi }{4}$.
The spin Chern numbers $\nu $ derived for $\theta = 0$ and $\theta = \tfrac{\pi }{4}$ are displayed in Figs. \ref{fig:3} (c) and \ref{fig:3} (d), respectively.
Note that $\nu $ is formulated by assuming two spin sectors, and is ill-defined for ${\Delta _s} = 0$.
This is the reason why $\nu $ is not quantized and widely fluctuates for $\theta = 0$.
In contrast, $\nu $ is exactly quantized to be integers in the entire region for $\theta = \tfrac{\pi }{4}$.
The angle $\theta = \tfrac{\pi }{4}$ is not special.
For an arbitrary $\theta $ in the range $\abs{\theta } \in (0,\tfrac{\pi }{2})$, ${\Delta _s}$ remains nonzero and $\nu $ is identically quantized (not shown).
An important implication of these results is that topological phases can be validly elucidated in terms of $\nu $ by using a proper operator ${s_\theta }$ even in the absence of spin conservation.

\begin{figure}
\centering
\includegraphics{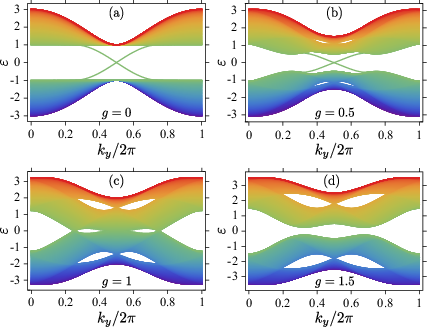}
\caption{Energy eigenvalues of $\mathcal{H}({k_y})$ for (a) $g = 0$, (b) $g = 0.5$, (c) $g = 1$, and (d) $g = 1.5$. The parameters used in the calculation are $(t,m) = (1,1)$ and ${N_x} = 100$.}
\label{fig:4}
\end{figure}

We have also investigated the energy spectra of the BHZ model in a cylindrical geometry with open boundaries at two lateral edges $x = \pm \tfrac{{{N_x}}}{2}$.
For this model, a 1D lattice Hamiltonian $\mathcal{H}({k_y})$ indexed by ${k_y}$ is derived after performing partial Fourier transformation.
The energy eigenvalues of $\mathcal{H}({k_y})$ are summarized in Fig. \ref{fig:4} for various coupling strengths $g$.
For $g < 1$, the helical edge modes traverse a finite energy gap.
As $g$ increases, the energy gap is reduced and eventually closed at $g = 1$.
For $g > 1.2$, the energy gap reopens, whereas the edge modes disappear.
These behaviors corroborate the above argument for bulk topology in view of the bulk-boundary correspondence.

\begin{figure}
\centering
\includegraphics{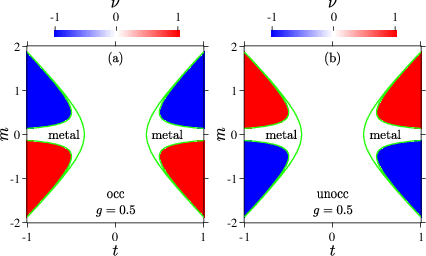}
\caption{Phase diagrams of momentum-space spin Chern number $\nu $ in the parameter space $(t,m)$ for (a) occupied and (b) unoccupied states. The parameters used in the calculation are $(g,\theta ) = (0.5,\tfrac{\pi }{4})$ and $M = 100$. The solid lines inserted in the figure represent the phase boundaries expected from the energy-gap closure. `occ' and `unocc' denote the occupied and unoccupied states, respectively.}
\label{fig:5}
\end{figure}

Figure \ref{fig:5} shows the phase diagram of $\nu $ derived for $g = 0.5$ and $\theta = \tfrac{\pi }{4}$.
As shown in the figure, the phase boundaries between quantized domains are appreciably deformed by the spin mixing term.
As expected, the nontrivial and trivial insulating phases are separated by the metallic gapless phases.
Note that the metallic phase is a topologically trivial phase.
As a consequence, $\nu $ always changes by $\pm 1$ at the boundaries.
In this sense, the spin mixing reconciles the phase boundaries of $\nu $ and ${\nu _{\mathbb{Z}_2}}$.

\section{Spin Chern number in momentum space}
\label{sec:4}

The spin Chern number argued above can be derived by discretizing the 2D BZ.
This approach is analogous to the one for the lattice Chern number \cite{ref:37}.
Here, we consider the lattice points ${{\mathbf{k}}_{\bm\ell} } = {\sum _{\mu}}{\ell _\mu }{{\mathbf{d}}_\mu }$ on the 2D BZ, where ${\bm\ell} = ({\ell _x},{\ell _y})$, ${\ell _\mu } \in \{ 0,1, \ldots ,{M_\mu } - 1\} $, and ${{\mathbf{d}}_\mu } = \tfrac{{2\pi }}{{{M_\mu }}}{{\mathbf{e}}_\mu }$.
The four corners of a square plaquette ${\bm\ell}$ of the BZ mesh are chosen as ${\mathbf{k}}_{\bm\ell} ^0 = {{\mathbf{k}}_{\bm\ell} }$, ${\mathbf{k}}_{\bm\ell} ^1 = {{\mathbf{k}}_{\bm\ell} } + {{\mathbf{d}}_x}$, ${\mathbf{k}}_{\bm\ell} ^2 = {{\mathbf{k}}_{\bm\ell} } + {{\mathbf{d}}_x} + {{\mathbf{d}}_y}$, and ${\mathbf{k}}_{\bm\ell} ^3 = {{\mathbf{k}}_{\bm\ell} } + {{\mathbf{d}}_y}$.
The local Wilson loop enclosing the plaquette ${\bm\ell}$ in each spin sector is given by
\begin{equation}
\label{eq:13}
{Q_\sigma }({\bm\ell}) = q_\sigma ^{03}({\bm\ell})q_\sigma ^{32}({\bm\ell})q_\sigma ^{21}({\bm\ell})q_\sigma ^{10}({\bm\ell}) ,
\end{equation}
where $q_\sigma ^{jj'}({\bm\ell}) = U_\sigma ^\dag ({\mathbf{k}}_{\bm\ell} ^j){U_\sigma }({\mathbf{k}}_{\bm\ell} ^{j'})$ is the overlap matrix between ${\mathbf{k}}_{\bm\ell} ^j$ and ${\mathbf{k}}_{\bm\ell} ^{j'}$.
In the limit of ${M_\mu } \to \infty $, $q_\sigma ^{jj'}({\bm\ell}) = {e^{i({\mathbf{k}}_{\bm\ell} ^j - {\mathbf{k}}_{\bm\ell} ^{j'}) \cdot {{\mathbf{A}}_\sigma }({\mathbf{k}}_{\bm\ell} ^{j'})}}$.
Hence,
\begin{equation}
\label{eq:14}
{Q_\sigma }({\bm\ell}) = \mathcal{P}{e^{i\oint_{\mathcal{C}({\bm\ell})} {d{\mathbf{k}} \cdot {{\mathbf{A}}_\sigma }({\mathbf{k}})} }} ,
\end{equation}
where $\mathcal{C}({\bm\ell})$ denotes the closed path joining ${\{ {\mathbf{k}}_{\bm\ell} ^j\} _{j = 0,1,2,3}}$.
Thus, the Chern number is expressed in each spin sector as
\begin{equation}
\label{eq:15}
\begin{split}
{C_\sigma } &= \frac{1}{{2\pi }} \Tr\sum\limits_{\bm\ell} {\oint_{\mathcal{C}({\bm\ell})} {d{\mathbf{k}} \cdot {{\mathbf{A}}_\sigma }({\mathbf{k}})} } \\
 &= \frac{1}{{2\pi }}\sum\limits_{p,{\bm\ell}} { \arg {\lambda _{p\sigma }}({\bm\ell})} ,\\
\end{split}
\end{equation}
where ${\lambda _{p\sigma }}({\bm\ell})$ is the eigenvalue of ${Q_\sigma }({\bm\ell})$.
The spin Chern number $\nu $ is derived from ${C_\sigma }$ by using Eq. (\ref{eq:11}).
The Wilson-loop formulation is applicable to generic time-reversal invariant 2D systems.
In the case of the BHZ model, ${Q_\sigma }({\bm\ell})$ is a scalar so that ${C_\sigma } = \tfrac{1}{{2\pi }}{\sum _{\bm\ell} } \arg {Q_\sigma }({\bm\ell})$.

We have numerically investigated the phase diagram of the spin Chern number following the above formulation.
As expected, the numerical results for the BHZ model are identical to those shown in Figs. \ref{fig:2} and \ref{fig:5}.
The numerical identity supports the mathematical equivalence of the two approaches described in Secs. \ref{sec:3} and \ref{sec:4}.
	
\section{Spin Chern number in real space}
\label{sec:5}

Thus far, the spin Chern number has been derived from a Wilson loop in momentum space.
The momentum-space formulation is, however, insufficient to elucidate topological phases in inhomogeneous or disordered systems that lack translational invariance.
In what follows, we generalize the theoretical treatment to derive a real-space formulation in a supercell framework, which can deal with ordered and disordered systems on an equal footing.

Let the eigenequation in real space be $\mathcal{H} \ket{u_n} = {\varepsilon _n} \ket{u_n} $.
The spin matrix is written in real space as
\begin{equation}
\label{eq:16}
S = {U^\dag }{s_\theta }U ,
\end{equation}
where ${U_{{\mathbf{r}}\alpha ,n}} = \braket{{\mathbf{r}},\alpha}{u_n} $ is the rectangular matrix composed of occupied (or unoccupied) eigenstates and ${\mathbf{r}} = ({l_x},{l_y})$ denotes the site position.
Diagonalizing $S$, we derive the spin eigenvalues ${s_i}$ and the spin eigenvectors $\ket{{v_i}} $.
The eigenvectors $\ket{v_{i\sigma }} $ in each spin sector are distinguished by $\sigma = \operatorname{sgn} ({s_i})$.
The spin eigenvector is expanded in terms of the energy eigenstates as
\begin{equation}
\label{eq:17}
\ket {u_{i\sigma }} = \sum\limits_n {\ket{u_n} \hspace{-3pt} \braket{u_n}{v_{i\sigma}}} .
\end{equation}
In matrix notation, this relation is expressed as ${U_\sigma } = U{V_\sigma }$, where $({U_{\sigma }})_{\mathbf{r}\alpha ,i} = \braket {{\mathbf{r}},\alpha }{{u_{i\sigma }}} $ and $({V_\sigma })_{ni} = \braket{{u_n}}{v_{i\sigma}}$.
These treatments are formally similar to the ones performed in momentum space.

In a supercell framework, we usually suppose a fictitious periodic system consisting of a large unit cell.
Assuming a supercell of size ${N_x} \times {N_y}$, the primitive translation vector of the reciprocal lattice is given by ${{\mathbf{b}}_\mu } = \tfrac{{2\pi }}{{{N_\mu }}}{{\mathbf{e}}_\mu }$.
The four corners of the BZ are at ${{\mathbf{k}}_0} = 0$, ${{\mathbf{k}}_1} = {{\mathbf{b}}_x}$, ${{\mathbf{k}}_2} = {{\mathbf{b}}_x} + {{\mathbf{b}}_y}$, and ${{\mathbf{k}}_3} = {{\mathbf{b}}_y}$.
Note that the spin matrix remains unchanged at these points due to the periodicity in the BZ.
The overlap matrix between two corners is reduced to
\begin{equation}
\label{eq:18}
\begin{split}
{(q_\sigma ^{jj'})_{ii'}} &= \braket{{u_{i\sigma }}({{\mathbf{k}}_j})}{{u_{i'\sigma }}({{\mathbf{k}}_{j'}})} \\
 &= \mel{u_{i\sigma }}{e^{i({{\mathbf{k}}_j} - {{\mathbf{k}}_{j'}}) \cdot {\mathbf{r}}}}{u_{i'\sigma }} \\
\end{split}
\end{equation}
with the help of the momentum translation operator ${e^{i({{\mathbf{k}}_j} - {{\mathbf{k}}_{j'}}) \cdot {\mathbf{r}}}}$ \cite{ref:33}.
In matrix notation, Eq. (\ref{eq:18}) is rewritten as $q_\sigma ^{jj'} = U_\sigma ^\dag {e^{i({{\mathbf{k}}_j} - {{\mathbf{k}}_{j'}}) \cdot {\mathbf{r}}}}{U_\sigma }$.
The path-ordered product expressed as
\begin{equation}
\label{eq:19}
{Q_\sigma } = q_\sigma ^{03}q_\sigma ^{32}q_\sigma ^{21}q_\sigma ^{10}
\end{equation}
forms the Wilson loop around the BZ boundary.
In the limit of ${N_\mu } \to \infty $, $q_\sigma ^{jj'} = {e^{i({{\mathbf{k}}_j} - {{\mathbf{k}}_{j'}}) \cdot {{\mathbf{A}}_\sigma }({{\mathbf{k}}_{j'}})}}$.
Hence,
\begin{equation}
\label{eq:20}
{Q_\sigma } = \mathcal{P}{e^{i\oint_{\partial {\text{BZ}}} {d{\mathbf{k}} \cdot {{\mathbf{A}}_\sigma }({\mathbf{k}})}}} .
\end{equation}
Therefore, the real-space Chern number is formulated in each spin sector as
\begin{equation}
\label{eq:21}
{C_\sigma } = \frac{1}{{2\pi }}\sum\limits_p {\arg {\lambda _{p\sigma }}} ,
\end{equation}
where ${\lambda _{p\sigma }}$ are the eigenvalues of ${Q_\sigma }$.
It is easy to show that ${C_\sigma }$ is quantized to be an integer for a large enough supercell \cite{ref:33}.
The spin Chern number $\nu $ is derived from ${C_\sigma }$ by using Eq. (\ref{eq:11}).

\begin{figure}
\centering
\includegraphics{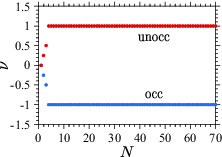}
\caption{Real-space spin Chern number $\nu $ as a function of system size $N$ in the absence of disorder. The parameters used in the calculation are $(t,m) = (1,1)$ and $g = \theta = 0$. `occ' and `unocc' denote the occupied and unoccupied states, respectively.}
\label{fig:6}
\end{figure}

First, we explore the real-space spin Chern number $\nu $ for a periodic ordered system of size $N = {N_x} = {N_y}$.
Figure \ref{fig:6} shows the size dependence of $\nu $ computed for $(t,m,g) = (1,1,0)$.
It is shown in the figure that the real-space formula correctly evaluates $\nu $ for moderately large systems.
The numerical accuracy is typically in the order of ${10^{-14}}$.
Note that the bulk invariant is derivable for $N \ge 4$, indicating that it is actually unnecessary to approach the thermodynamic limit $N \to \infty $.
This observation is similar to the one made for the real-space Chern number argued in a previous study \cite{ref:33}, and does not contradict Eq. (\ref{eq:21}).
It is also confirmed that the phase diagrams in the parameter space $(t,m)$ are identical for the real and momentum-space formulations irrespective of $g$.

\begin{figure}
\centering
\includegraphics{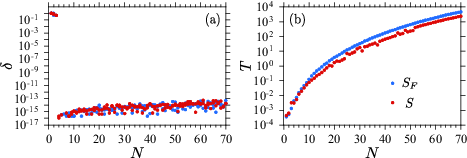}
\caption{Numerical evaluation of the two approaches using $S$ and ${S_F}$. Panel (a) shows the absolute error $\delta $ and panel (b) shows the computation time $T$ as a function of system size $N$ in the absence of disorder. The parameters used in the calculation are $(t,m) = (1,1)$ and $g = \theta = 0$.}
\label{fig:7}
\end{figure}

It may be appropriate here to compare the present formulation with the ones carried out for the spin Bott index \cite{ref:26,ref:27} and the related methods \cite{ref:24,ref:25,ref:28,ref:29,ref:30}.
In the previous methods, the spin operator projected onto occupied states ${S_F} = F{s_z}F$ is used to define two spin sectors, with $F = U{U^\dag }$ denoting the Fermi projector.
This operator can be generalized as ${S_F} = F{s_\theta }F$.
It is easy to show that $\ket {u_{i\sigma }} $ defined by Eq. (\ref{eq:17}) satisfies ${S_F} \ket{u_{i\sigma }} = {s_{i\sigma }} \ket{u_{i\sigma }} $.
This means that the spin eigenvectors are the same in the present and previous formulations.
It is also shown in a previous study \cite{ref:33} that the Sylvester's determinant identity reduces the Bott-index formula to the Wilson-loop formula.
Thus, the two approaches using different matrices, $S$ and ${S_F}$, are mathematically equivalent.
However, there is a large difference in the matrix size.
For the BHZ model, the size of $S$ is $L \times L$, whereas it is $2L \times 2L$ for ${S_F}$, with $L = 2{N_x}{N_y}$ being the number of occupied (or unoccupied) states.
Clearly, ${S_F}$ has many surplus eigenvalues.
Since the extra eigenvalues are all zero, they are not physically meaningful \cite{ref:27}.
Diagonalizing the unnecessarily large matrix and eliminating the irrelevant eigenvectors tend to prevent an efficient computation.
To show this quantitatively, we have compared the two approaches based on $S$ and ${S_F}$ in terms of computational time.
The numerical results of this comparison are summarized in Fig. \ref{fig:7}.
As demonstrated in this figure, there is an appreciable difference in computational time for large systems.
Quantitatively, the present approach enables 2-4 times faster computation.

Finally, we address a disordered BHZ model.
A nonmagnetic random potential can be expressed as ${\mathcal{H}_2} = {s_0} \otimes {H_2}$, where
\begin{equation}
\label{eq:22}
{H_2} = \sum\limits_{{\mathbf{r}},\tau } \ket{{\mathbf{r}},\tau } {u_{{\mathbf{r}}\tau }} \bra{{\mathbf{r}},\tau }
\end{equation}
and $\tau $ denotes the orbital degrees of freedom.
We regard ${u_{{\mathbf{r}}\tau }} \in [-W,W]$ as a random variable following the uniform distribution.
The disordered BHZ model is described as $\mathcal{H} = {\mathcal{H}_0} + {\mathcal{H}_1} + {\mathcal{H}_2}$.

\begin{figure}
\centering
\includegraphics{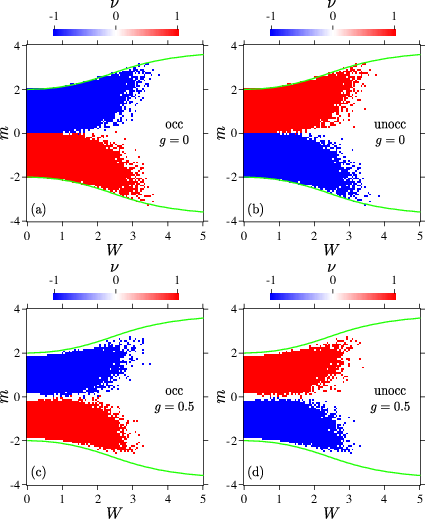}
\caption{Real-space spin Chern number $\nu $ as a function of $W$ and $m$ for [(a) and (b)] $(g,\theta ) = (0,0)$ and [(c) and (d)] $(g,\theta ) = (0.5,\tfrac{\pi }{4})$. The left panels, (a) and (c), and the right panels, (b) and (d), show the numerical results for occupied and unoccupied states, respectively. The parameters used in the calculation are $t = 1$ and $N = 20$. The solid lines inserted in the figure represent the phase boundaries predicted from Eqs. (\ref{eq:23}) and (\ref{eq:24}). `occ' and `unocc' denote the occupied and unoccupied states, respectively.}
\label{fig:8}
\end{figure}

Figures \ref{fig:8} (a) and \ref{fig:8} (b) show the real-space spin Chern number $\nu $ as a function of $W$ and $m$ for $t=1$ and $g=0$.
Each data point corresponds to a single realization of the random potential.
In the weak disorder limit $W \to 0$, $\abs{\nu} = 1$ for $\abs{m} < 2$ and 0 for $\abs{m} > 2$.
The phase boundary separating the nontrivial and trivial regimes varies with increasing $W$.
For strong enough disorder, the system is eventually in the trivial phase irrespective of $m$.
A topological Anderson insulator phase \cite{ref:38} is seen for $\abs{m}$ just exceeding 2.
In this region, weak disorder induces the trivial-to-nontrivial phase transition.

It is well known that the topological Anderson insulator phase is accounted for by the self-consistent Born approximation \cite{ref:39,ref:40,ref:41}.
In this approximation, the retarded self-energy matrix is represented as $\Sigma = {s_0} \otimes ({\Sigma _0}{\tau _0} + {\Sigma _z}{\tau _z})$ for $g=0$.
The two scalar factors, ${\Sigma _0}$ and ${\Sigma _z}$, satisfy the self-consistent equations
\begin{equation}
\label{eq:23}
{\Sigma _0} = \frac{{{W^2}}}{{3{N_x}{N_y}}}\sum\limits_{\mathbf{k}} {\frac{{\varepsilon - {\Sigma _0}}}{{{D_{\mathbf{k}}}}}} ,
\end{equation}
\begin{equation}
\label{eq:24}
{\Sigma _z} = \frac{{{W^2}}}{{3{N_x}{N_y}}}\sum\limits_{\mathbf{k}} {\frac{{{h_z} + {\Sigma _z}}}{{{D_{\mathbf{k}}}}}} ,
\end{equation}
where ${D_{\mathbf{k}}} = {(\varepsilon - {\Sigma _0})^2} - (h_x^2 + h_y^2) - {({h_z} + {\Sigma _z})^2}$.
The solution of Eqs. (\ref{eq:23}) and (\ref{eq:24}) replaces the bare parameter $m$ with the renormalized one $\tilde m = m + \Re \Sigma _z$.
Note that $\Re \Sigma _0$ corresponds to merely a constant potential for all bands and is irrelevant to topological phase transitions.
As shown in Figs. \ref{fig:8} (a) and \ref{fig:8} (b), the weak disorder boundaries for $g=0$ are well described in terms of the renormalized parameter $\tilde m$.
	
Figures \ref{fig:8} (c) and \ref{fig:8} (d) explain the topological phases for $g=0.5$ in the presence of disorder.
In the weak disorder limit, the trivial metallic phase emerges in the vicinity of $m=0$.
A finite $g$ also modifies the upper and lower phase boundaries.
These features do not contradict the results shown in Fig. \ref{fig:5}.
The phase boundaries vary as $W$ increases.
The $W$-dependence observed for $g=0.5$ does not largely differ from the one observed for $g=0$.
This implies that the topological Anderson insulator phase persists even in the absence of spin conservation as long as $g$ is small enough.

\section{Summary}
\label{sec:6}

We have formulated the spin polarization and the spin Chern number in momentum and real spaces in terms of the Wilson loop in each spin sector.
These formulations employ the spin matrix in the occupied or unoccupied subspace.
In this study, the spin matrix is generalized by the unitary rotation to prevent the spectral gap from closing.
In the presence of a finite spin gap, the spin eigenvectors reconstruct the eigenstates of the Hamiltonian in two spin sectors.
	
The charge polarization in each spin sector is formulated in terms of the Wilson loop along the effective 1D BZ created by a dimensional reduction.
A linear combination of the charge polarizations defines the spin polarization and naturally quantifies the spin Chern number in momentum space.
It is also shown that the spin Chern number is directly derived from the local Wilson loops on a discrete 2D BZ.
	
The theoretical treatment is further generalized for real space in a supercell framework, which can deal with ordered and disordered systems on an equal footing.
In this framework, the spin Chern number in real space is formulated in terms of the Wilson loop around the BZ boundary.
For large enough supercells, the real-space spin Chern number is quantized at integers.
The spin Chern number derived in this way mathematically coincides with the spin Bott index, but is greatly simplified and makes numerical computations more efficient.
	
The Wilson-loop formulations were used to numerically examine topological phases in the ordered and disordered BHZ models in the presence or absence of spin mixing.
Numerical results show an equivalence of the spin Chern number and the ${\mathbb{Z}_2}$ index.
Particularly, it is found that spin mixing induces the metallic phase, which is a topologically trivial phase in view of the spin Chern number.
It is also demonstrated that the topological Anderson insulator phase generated by disorder persists even in the presence of spin mixing as long as the mixing is weak enough.

\bibliography{ref}
 
\end{document}